\documentclass[a4paper]{jpconf}

\usepackage{graphics}

\usepackage{amsfonts}
\usepackage{amsmath}
\usepackage{amssymb}
\usepackage{upgreek}
\usepackage{bm}
\usepackage{dcolumn}
\usepackage{epsfig}
\usepackage{graphicx}

\usepackage[latin1]{inputenc}
\usepackage{latexsym}
\usepackage{rotating}
\usepackage{hyperref}

\newcommand\be{\begin{equation}}
\newcommand\ba{\begin{eqnarray}}
\newcommand\ee{\end{equation}}
\newcommand\ea{\end{eqnarray}}

\newcommand{\MAT}{{\mbox{\tiny mat}}}

\newcommand{\MIN}{{\mbox{\tiny min}}}
\newcommand{\INC}{{\mbox{\tiny inc}}}
\newcommand{\APO}{{\mbox{\tiny apo}}}
\newcommand{\PERI}{{\mbox{\tiny peri}}}

\newcommand{\met}{\mbox{g}}

\begin{document}

\title{Testing Chern-Simons modified gravity with observations of extreme-mass-ratio binaries}

\author{P.~Canizares$^{1}$, J.R.~Gair$^{2}$ and C.F.~Sopuerta$^{3}$}
\address{$^{1,3}$Institut de Ci\`encies de l'Espai (CSIC-IEEC), 
Facultat de Ci\`encies, Campus UAB, Torre C5 parells, 
Bellaterra, 08193 Barcelona, Spain.\\ 
$^{1,2}$Institute of Astronomy, Madingley Road, Cambridge, CB30HA, United Kingdom.}

\ead{$^{1}$pcm@ast.cam.ac.uk,$^{2}$jrg23@cam.ac.uk, $^{3}$sopuerta@ieec.uab.es}

%----- ABSTRACT -----%
\begin{abstract} 
Extreme-Mass-Ratio Inspirals (EMRIs) are one of the most promising sources of gravitational waves 
(GWs) for space-based detectors like the Laser Interferometer Space Antenna (LISA). EMRIs consist of a
compact stellar object orbiting around a massive black hole (MBH). Since EMRI signals are expected to 
be long lasting (containing of the order of hundred thousand cycles), they will encode the structure 
of the MBH gravitational potential in a precise way such that features depending on the theory of 
gravity governing the system may be distinguished.  That is, EMRI signals may be used to test gravity 
and the geometry of black holes. However, the development of a practical methodology for computing
the generation and propagation of GWs from EMRIs in theories of gravity different than
General Relativity (GR) has only recently begun. In this paper, we present a
parameter estimation study of EMRIs in a particular modification of GR, which is described
by a four-dimensional Chern-Simons (CS) gravitational term. We focus on
determining to what extent a space-based GW observatory like LISA could
distinguish between GR and CS gravity through the detection of GWs from EMRIs.
\end{abstract}

%----- INTRODUCTION -----%
\section{Introduction}
\label{intro}
It is known that surrounding the central MBH of many quiescent galaxies there are around $10^7 - 10^8$ stars 
forming a cusp/core (see e.g.~\cite{Amaro2011}). As a consequence of mass segregation and large scattering 
encounters between stars in the cluster, the largest stellar-mass compact objects (SCOs), like neutron stars, 
$m\approx 1.4 M_{\odot}$, white dwarfs, $m \approx 0.6 M_{\odot}$, and stellar-mass black holes, 
$m \approx 1-50 M_{\odot}$, may be perturbed onto an orbit that passes close enough to the central MBH and become gravitationally bound to it. It is expected that the capture of a SCO by a MBH may be 
a frequent phenomenon in the Universe. The MBHs of interest for a GW observatory like LISA~\cite{LISA} have
masses in the range $M_{\bullet}\sim 10^5-10^7 M_{\odot}$.  In addition, there are observations suggesting 
the existence of Intermediate-Mass Black Holes (IMBHs), whose masses are in the range 
$M_{\bullet}\sim 10^2-10^4 M_{\odot}$ (see e.g.~\cite{AmaroSeoane:2007aw}). These may form EMRI-like systems, 
either an IMBH-MBH inspiral or an SCO-IMBH system. Space-based detectors like LISA will be sensitive to the 
inspiral of an IMBH into a MBH, whereas ground-based detectors like the Einstein Telescope~\cite{ET} will be sensitive 
to the inspiral of SCOs into IMBHs.

Once the SCO is captured, the binary loses energy and angular momentum through the emission of GWs. This drives 
the inspiral of the SCO until it plunges into the MBH. Due to the fact that the mass of the SCO is much smaller 
than the mass of the MBH, with typical mass ratios lying in the range $\mu=m/M_{\bullet} \sim 10^{-7}-10^{-3}$, 
the inspiral can be modelled as a slow adiabatic process in which the orbital time scale is much shorter than the 
timescale on which the orbit shrinks. 

During the last year of evolution before plunge, the orbit of the SCO spends of the order of $10^{5}$ cycles
in the strong field region of the MBH, mapping its geometric structure (which can be characterized
in terms of mass and current multipolar moments).  This map is encoded in the shape and timing of the GWs
emitted~\cite{Ryan52}. If we can decode this map, EMRI detections will allow us to study different aspects of 
the MBHs located at the centers of galaxies, the stellar dynamics around galactic centers, etc.  In particular,
EMRI GW signals may be very useful to test the spacetime geometry of 
MBHs~\cite{Collins:2004ex,Glampedakis:2005cf,Barack:2007p,Yunes:2008gb,Brink:2008xx,Vigeland:2009pr,Hughes:2010xf,Sopuerta:2010zy} 
and even alternative theories of gravity (see, e.g.~\cite{Hughes:2006pm,Schutz:2009zz,Sopuerta:2010zy,Babak:2010ej,Vigeland:2011ji,Gair:2011ym}),
which is the subject of this article.

Due to the myriad  of alternative theories of gravity available, the question that arises here is which kind of 
theory do we choose to test? Alternatively, what kind of phenomena can we expect to test using EMRI observations? 
In this paper, we will focus on Dynamical Chern-Simons Modified Gravity (DCSMG) which is a 4D effective theory of 
gravity that arises in certain low-energy limits of \emph{string theory} and can also arise in
\emph{Loop Quantum Gravity} (see~\cite{Alexander:2009} for a review of DCSMG). DCSMG has been proposed as an 
explanation for the cosmic baryon asymmetry~\cite{Alexander:2004us} and makes predictions for the polarisation of 
the cosmic microwave background~\cite{Lue:1998mq}. In addition, it has potentially observable signatures in EMRI
GW observations~\cite{Sopuerta:2009iy}.

In this paper we summarize the necessary techniques that we have developed to estimate the capability of a LISA-like GW detector to
distinguish between GR and DCSMG using GW signals from EMRIs, extending and improving the results of~\cite{Sopuerta:2009iy}. 
A more detailed account of these developments will appear elsewhere~\cite{cgs:2012}.
Due to the large difference between the masses of the components in the binary, an EMRI signal can be modelled 
accurately using black hole perturbation theory (see e.g.~\cite{Drasco:2005kz}), in which the SCO is represented 
as a particle orbiting in a given MBH spacetime background. In this framework, at the lowest order of approximation
the orbit of the particle is a geodesic of the MBH spacetime, but radiation reaction (RR) effects, i.e.~the 
effects that arise from the interaction of the SCO with its own gravitational field, accelerate the trajectory 
and then drive the inspiral of the SCO.  Depending on the theory of gravity that describes the system, the RR effects drive the inspiral in different ways . 

This paper is organised as follows. In Sec.~\ref{formulation} we summarize the formulation of our EMRI model in DCSMG. 
In Sec.~\ref{signal_analysis} we review the basics of signal analysis employed in our study. Finally, in Sec.~\ref{Results}, 
we present results concerning whether a GW detector like LISA will be able to discriminate between an EMRI in GR and
one in DCSMG. 

Throughout this paper, we follow the conventions of Misner, Thorne and Wheeler~\cite{Misner:1973cw}:
Greek letters will represent spacetime indexes and Latin letters in the middle of the alphabet $i,j,...,$ will represent 
spatial indexes only. Partial derivatives are denoted by $\partial_{\alpha}h=\partial h/\partial x^{\alpha}=h_{, \alpha}$,
while, for any quantity $h$, covariant derivatives are denoted by $\nabla_{\alpha} h=h_{;\alpha}$. In addition, we use 
the Einstein summation convention and geometrised units, i.e.~$G=c=1$ and the gravitational constant is given by 
$\kappa = 1/(16\pi G)$. Finally, we use a bar, e.g.~$\bar{h}$, to denote background quantities.

%---- OUR MODEL FORMULATION AND ITS EVOLUTION----%
\section{Formulation of our EMRI model} \label{formulation}
We approximate our system following the procedure described in~\cite{Sopuerta:2009iy}, in which an EMRI is modelled as a 
binary composed of a point particle moving in a modified Kerr MBH background geometry. That is, the MBH geometry is slightly modified as a consequence 
of the CS gravitational modifications to GR. The motion of the particle, at the lowest order of approximation, is given 
by a geodesic of this CS-modified MBH geometry. The action describing the system is given by:
\begin{eqnarray}
S = \kappa S_{EH} + \frac{\alpha}{4}S_{CS} +\frac{\beta}{2} S_{\vartheta} +S_{\MAT}\,,\label{action}
\end{eqnarray}
where the different terms are:
\begin{eqnarray}
S_{EH}&=&\int d^4x\sqrt{-\met}R\;, \label{EH}\\
S_{CS}&=& \int d^4x\sqrt{-\met}\,\vartheta\;{}^{\ast}RR\;,
\label{CScorrection}\\
S_{\vartheta}&=& -\int d^4x\sqrt{-\met}\,\left[\ \met^{\mu\nu}(\nabla_{\mu}\vartheta)(\nabla_{\nu}\vartheta)\right]\;, \label{CSaction}\\
S_{\MAT}&=&\int d^4\sqrt{-\met} \ {\cal L}_{\MAT}\,,\label{Matteraction}
\end{eqnarray}
which represent the Einstein-Hilbert action [Eq.~(\ref{EH})], the CS gravitational correction [Eq.~(\ref{CScorrection})], 
the CS scalar field action [Eq.~(\ref{CSaction})], and the action of the matter degrees of freedom [Eq.~(\ref{Matteraction})].  
The coupling constant, $\alpha$, is associated with the Pontryagin density, 
${}^{\ast}RR:=\frac{1}{2}\epsilon^{\gamma\delta\mu\nu}R^{\alpha}_{\\beta\mu\nu}R^{\beta}_{\ \alpha\gamma \delta}$, which is 
coupled to the CS scalar field, $\vartheta$. The coupling constant, $\beta$, determines the strength of the coupling of the CS 
scalar field with the stress-energy distribution. 

Following~\cite{Yunes:2009hc}, we can construct the modified MBH geometry and the CS scalar field using the slow-rotation ($a/M_{\bullet}\ll 1$, 
where $M_{\bullet}$ and $a=|\bf{S}/M_{\bullet}|$ are the MBH mass and spin respectively) and weak-coupling approximation. Then, the 
spacetime metric, $\met_{\mu\nu}$, and the CS scalar field, $\vartheta$, can be written as the sum of a background and a perturbative 
quantity characterized by the perturbative parameter $\upsilon$:
\begin{eqnarray}
\met_{\mu\nu}&=& \bar{\met}_{\mu\nu}+\upsilon h_{\mu\nu}+ {\cal O}(\upsilon^2) \,,  \\
\vartheta &=& \bar{\vartheta}+\upsilon\vartheta + {\cal O}(\upsilon^2)\,.
\end{eqnarray}
The MBH metric has the following non-zero components in Boyer-Lindquist-like  coordinates $(t,r,\theta,\phi)$~\cite{Yunes:2009hc}:
\begin{eqnarray}
&&\bar{\met}^{}_{tt} = -\left(1-2M_{\bullet}\frac{r}{\rho^2}\right)\,,\label{DCSMG_metric_0}\\
&&\bar{\met}^{}_{rr} = \frac{\rho^2}{\Delta}\,,\\
&&\bar{\met}^{}_{\theta \theta}=\rho^2 \,,\\
&&\bar{\met}^{}_{\phi \phi}=\frac{\Sigma}{\rho^2}\sin^2{\theta}\,,\\
&&\bar{\met}^{}_{t\phi} =\left[\ \frac{5}{8}\frac{\xi}{M_{\bullet}^{4} } \frac{M^5_{\bullet}}{r^4}\left(1+\frac{12M_{\bullet}}{7r} 
+ \frac{27M^2_{\bullet}}{10r^{2}}\right) - 2M^2_{\bullet}\frac{r}{\rho^2}\ \right]\frac{a}{M_{\bullet}}\sin^2\theta\;\,,
\label{DCSMG_metric}
\end{eqnarray}
where $\Sigma=(r^2+a^2)^2-a^2\Delta\sin^2\theta$, $\rho^2=r^2+a^2\cos^2\theta$ and $\Delta = r^2f+a^2$, with $f = 1-2M_{\bullet}/r$. 
The CS parameter $\xi$ is given by
\begin{eqnarray}
\xi:=\frac{\alpha^2}{\beta\kappa}\,.\label{xi}
\end{eqnarray}
Notice that the only metric component that contains the CS parameter $\xi$ is $\bar{\met}_{t\phi}$ and that it behaves as $r^{-4}$ 
for $M_{\bullet}/r\ll 1$, decaying faster than the rest of metric contributions and becoming negligible at large distances. Under the 
preceding approximations, the CS scalar field is given by~\cite{Yunes:2009hc}:
\begin{eqnarray}
\bar{\vartheta}=\frac{5}{8}\frac{\alpha}{\beta}\frac{a}{M_{\bullet}}\frac{\cos\theta}{r^2}\left(1 + \frac{2M_{\bullet}}{r} 
+ \frac{18M_{\bullet}^2}{5r^2} \right) \,. \label{CS_scalar_field}
\end{eqnarray}

The numerical evolution of the system comprises two stages, namely the computation of the SCO trajectory around the MBH and then 
the computation of the corresponding waveform based on this trajectory. To generate the trajectory, it has been shown 
in~\cite{Sopuerta:2009iy} that in the test-mass limit the SCO must follow geodesics of the CS deformed MBH background.
It turns out that the metric [Eqs.~(\ref{DCSMG_metric_0})-(\ref{DCSMG_metric})] has almost identical symmetry properties as the Kerr metric (at least at the
level of approximation used for its derivation).  As a consequence, the geodesic motion of the particle is obtained by solving 
the following set of Ordinary Differential Equations (ODEs):
\begin{eqnarray}
\dot{t}&=&\dot{t}_K + L_z\delta\met^{CS}_{\phi}\,,\label{geoCS_1} \\
\dot{\phi} &=&\dot{\phi}_K - E\delta\met^{CS}_{\phi}\,,\label{geoCS_1b}\\
\dot{r}^2 &=& \dot{r}^2_K + 2EL_zf\delta\met^{CS}_{\phi}\,, \label{geoCS_2}\\ 
\dot{\theta}^2& =& \dot{\theta}^2_K\,, \label{dottheta}
\end{eqnarray}
where $E$ is the energy and $L_z$ is the $z$-component of the orbital angular momentum of the particle per unit mass. Here,
a dot denotes differentiation with respect to the proper time and ($\dot{t}_K$, $\dot{r}_K$,$\dot{\theta}_K$, $\dot{\phi}_K$) 
are the corresponding expressions for geodesics in a Kerr MBH background:
\begin{eqnarray}
\rho^2\dot{t}_K &=& -\left(a^2E\sin^2\theta-aL_z\right) + \left(r^2+a^2\right)\left(\left(r^2+a^2\right)\frac{E}{\Delta}-
\frac{aL_z}{\Delta} \right)\,, \label{geoK_1}\\
\rho^2\dot{\phi}_K &=&  -\left(aE-\frac{L_z}{\sin^2\theta}\right) + (r^2+a^2)\frac{aE}{\Delta}-\frac{a^2L_z}{\Delta} 
\,,\label{geoK_1b} \\
\rho^4\dot{r}_K^2 &=& \left[(r^2 + a^2)E -aL_z\right]^2 -\Delta\left[ {\cal Q} + (aE - L_z)^2 +r^2\right]\,, \label{geoK_2}\\
\rho^4\dot{\theta}^2_K &=& {\cal Q} - \cot^2\theta L_z^2 - a^2\cos^2\theta(1 -
E^2)\,,
\end{eqnarray}
where ${\cal Q}$ denotes the Carter constant per unit mass squared. The correction $\delta\met^{CS}_{\phi}$ appearing in 
Eqs.~(\ref{geoCS_1})-(\ref{dottheta}) is the CS correction, which is given by:
\begin{eqnarray}
\delta\met^{CS}_{\phi}=\frac{\xi a}{112\,r^8f}\left(70r^2+120rM_{\bullet}+189M^2_{\bullet} \right)\,.\label{cs_correction}
\end{eqnarray}

In order to avoid the presence of turning points in the integration of these equations of motion, we introduce two 
new angle coordinates, $\psi$ and $\chi$, associated with the radial and polar motion respectively, which are defined 
by the equations:
\begin{eqnarray}
r = \frac{pM_{\bullet}}{1+e\cos\psi}\,,~~~~~\cos^2\theta = \cos^2\theta_{\MIN}\cos^2\chi\,, \label{new_coordinates}
\end{eqnarray}
where $p$ and $e$ are the semilatus rectum and the eccentricity of the orbit respectively, and $\theta_{\MIN}$ is the 
minimum value of the polar angle $\theta$ in the orbit.  This value is related to the angle of
inclination of the orbit, $\theta_{\INC}$ (see Table~\ref{parameters}), through
\begin{eqnarray}
\theta^{}_{\MIN}= \frac{\pi}{2}-\theta^{}_{\INC}\,.
\end{eqnarray}
A given orbit can be characterized in terms of the constants of motion $(E,L_{z},{\cal Q})$ or, alternatively, in terms
of the orbital parameters $(e,p,\theta_{\INC})$.   To transform between these two sets of constants we use the equations
at the turning points of the radial and polar coordinates.  
At $\theta = \theta_{\MIN}$ we have that $\dot{\theta}=0$, leading to an expression for ${\cal Q}$~\cite{Sopuerta:2009iy}:
\begin{eqnarray}
{\cal Q} = \cos^2\theta_{\MIN}\left[ \frac{L_z^2}{\sin^2\theta_{\MIN}} + a^2(1-E^2)\right]\,.\label{carter_point}
\end{eqnarray}
The additional two equations needed are those that define the radial turning points, that is, the pericenter, $r_{\PERI}$, and apocenter, $r_{\APO}$, which we relate to $(e,p)$ in the usual way:
\begin{eqnarray}
r^{}_{\PERI}=\frac{pM_{\bullet}}{1+e}\,,~~~~~~r^{}_{\APO}=\frac{pM_{\bullet}}{1-e}\,.
\end{eqnarray}
Eq.~(\ref{carter_point}) together with the equations $\dot{r}(r_{peri})=\dot{r}(r_{apo})=0$ allow  us to relate the values of the constants $(E,L_{z},{\cal Q})$ to those of the orbital parameters $(e,p,\theta_{\INC})$.

Until now we have only considered geodesic motion.  Constructing waveforms from geodesic trajectories can lead to a
a confusion problem in the sense that it is possible that an EMRI waveform in GR matches an EMRI waveform in DCSMG 
but with different parameters, given that the motion in both cases is going to be characterized by just three fundamental 
frequencies of motion.  This degeneracy is overcome when RR effects are taken into account, since these  tend to break such confusion, as the RR effects drive the inspiral in different ways depending on the theory of gravity that describes 
the EMRI system.  Nevertheless, it was shown in~\cite{Sopuerta:2009iy} that, at leading order, the GW emission in DCSMG 
takes the same form as in GR.  So, as a first approximation we describe RR effects in DCSMG using expressions from
GR.  In particular we use the fluxes for the constants of motion $(E,L_{z},{\cal Q})$ provided in~\cite{Gair2006}.
In principle one might think that this would lead to exactly the same inspiral evolution in DCSMG and GR, but it is not the
case due to the different relations between the constants of motion $(E,L_{z},{\cal Q})$ and the orbital parameters 
$(e,p,\theta_{\INC})$.

To compute the EMRI waveforms in DCSMG, we first solve the ODEs given in Eqs.~(\ref{geoCS_1})-(\ref{dottheta}) for the 
angle variables $(\psi(t),\chi(t),\varphi(t))$ (see e.g.~\cite{Drasco:2003ky}). The scheme for evolving a particular  
EMRI in DCSMG can be summarized as follows: For a given set of initial orbital parameters, 
$(e^{(0)},p^{(0)},\theta^{(0)}_{\INC})$, we find the associated initial constants of the motion, 
$(E^{(0)},L_z^{(0)},{\cal Q}^{(0)})$, which are different from the ones that one would obtain in GR for the same orbital
parameters.  Then, the time evolution of the ``constants'' of motion, $(\dot{E}^{(0)}, \dot{L}_z^{(0)},\dot{{\cal Q}}^{(0)})$,
the RR effect, is computed using the fluxes of~\cite{Gair2006}.  From the  values of the constants of motion, 
$(E^{(0)},L_z^{(0)},{\cal Q}^{(0)})$, their evolution due to RR, $(\dot{E}^{(0)}, \dot{L}_z^{(0)},\dot{{\cal Q}}^{(0)})$, 
and the value of the geodesic radial period, $T_r$  (the coordinate time to go from the apocenter to the pericenter and 
back again to apocenter), we obtain the values of the new constants of motion, $(E^{(1)},L_z^{(1)},{\cal Q}^{(1)})$. 
Finally, from the new values of the constants of motion we find the new values of the orbital parameters
$(p^{(1)},e^{(1)},\theta_{\INC}^{(1)})$. This algorithm is iterated along the whole EMRI evolution to obtain the 
SCO orbit around the MBH. The gravitational waveform is then generated using the multipolar formalism given by
Thorne~\cite{Thorne1980R} up to quadrupolar order, by identifying the Boyer-Lindquist coordinates with spherical polar 
coordinates in a pseudo flat-space.

%----- SIGNAL ANALYSIS FORMALISM -----%
\section{Signal analysis formalism}\label{signal_analysis}
In this section we briefly review the formalism of signal analysis that we use in this paper. For our calculations we 
use the LISA noise sensitivity curve (see e.g.~\cite{Barack:2003fp}) to represent a possible future space-based GW 
detector. At low-frequencies the response of the LISA detector can be decomposed as the response of a pair of independent, 
right-angle Michelson detectors, which we label by $\alpha=I,II$ in the standard way. Each of these data streams is a 
time series, $s_\alpha(t)$, which is a superposition of a possible gravitational wave signal $h^{}_{\alpha}(t)$ and
noise $n^{}_{\alpha}(t)$, $s^{}_{\alpha}(t) = h^{}_{\alpha}(t)+n^{}_{\alpha}(t)$.  Under the assumptions that the noise 
in the detector is stationary, Gaussian and uncorrelated between the two data streams, and that the spectral density of the
noise is the same in the two channels, the Fourier components of the noise, $\tilde{n}_\alpha(f)$, would have a Gaussian 
probability distribution and
\be
\langle\; \tilde{n}^{}_{\alpha}(f)\,\tilde{n}^{}_{\beta}(f')^* \;\rangle = \frac{1}{2} \,\delta^{}_{\alpha\beta}\,
\delta(f-f')\, S^{}_{n}(f)\,, 
\ee
where a tilde denotes the Fourier transform, an asterisk denotes complex conjugation, angle brackets denote the ensemble 
expectation value, and $S^{}_{n}(f)$ is the one-sided noise spectral density of the data stream, which is the same for 
both channels.

The vector space of signals has a natural inner product associated with $S^{}_n(f)$:
\be
({\bf a}|{\bf b}) = 2\,\sum^{}_{I,II} \int_0^\infty \frac{\tilde{a}^{*}_{\alpha}(f) \tilde{b}^{}_{\alpha}(f) 
+ \tilde{a}^{}_{\alpha}(f)\tilde{b}^{*}_{\alpha}(f)}{S^{}_n(f)} {\rm d}f \,.
\ee
The assumption of Gaussianity then means that the probability that the noise, ${\bf n}$, has a particular realisation, ${\bf n}_0$, 
has the form
\be
p({\bf n}={\bf n}_0) \propto e^{-({\bf n}_0|{\bf n}_0)/2} \,,\label{prob}
\ee
and hence the likelihood that a signal ${\bf h}$ is present in the data stream ${\bf s}$ is proportional to 
$\exp(-({\bf s}-{\bf h}|{\bf s}-{\bf h})/2)$. The `best-fit' waveform will be the one that maximises $({\bf s}|{\bf h})$ and,
thus, provides the maximum likelihood parameter estimate.  The expected signal-to-noise ratio (SNR) when filtering with 
the correct waveform is
\be
\text{SNR}=\frac{({\bf h}|{\bf h})}{{\rm rms}({\bf h}|{\bf n})} = \sqrt{({\bf h}|{\bf h})}\,,
\ee
where the second equality follows from the fact that the expectation value of $({\bf a}|{\bf n})({\bf b}|{\bf n})$ 
is $({\bf a}|{\bf b})$~\cite{Cutler:1994ys}.

A given template family ${\bf h}(t;\vec{\lambda})$ will depend on a set of parameters $\vec{\lambda}$. If we expand 
$\exp(-({\bf s}-{\bf h}|{\bf s}-{\bf h})/2)$ around the best-fit parameters, $\vec{\lambda}_0$, by writing
$\vec{\lambda} = \vec{\lambda}_0 + \delta \vec{\lambda}$, we can derive a probability distribution function for the 
parameter uncertainties
\be
p(\delta\vec{\lambda}) = {\cal N} \exp\left(-\frac{1}{2} \Gamma_{jk} \delta\lambda^j\delta\lambda^k\right)\,,
\label{likelihood}
\ee
where ${\cal N}=\sqrt{\det( \Gamma /2 \pi)}$ is the normalization factor and $\Gamma_{ij}$ is the 
\emph{Fisher information matrix}~\cite{Fisher:1935}
\be
\Gamma_{jk} =\left( \frac{\partial {\bf h}}{\partial \lambda^j} \vline\frac{\partial {\bf h}}{\partial \lambda^k} \right) 
\,.\label{FM}
\ee
The variance-covariance matrix for the waveform parameters is therefore given by the inverse of the Fisher matrix
\be
\langle \delta\lambda^j\delta\lambda^k \rangle = \left( \Gamma^{-1}\right)_{jk} + O(1/{\rm SNR})\label{covariance}
\ee
and so we can estimate the precision with which we will be able to measure a particular parameter, $\lambda_i$, 
using LISA observations by computing the component $\Gamma^{-1}_{ii}$ of this inverse matrix.

In our simulations, the space of EMRI waveforms in DCSMG, $\{h(t;\lambda^i)\,|\,i=1,\ldots,N\}$ is characterised by the 
following 15 parameters (see Table~\ref{parameters} for a description of these parameters): 
\begin{eqnarray}
\vec{\lambda} = \{M^{}_{\bullet}, a, \mu, e^{}_0, p^{}_0, \theta^{\INC}_0, \hat{\xi}, \theta^{}_S, \phi^{}_S, 
\theta^{}_K,  \phi^{}_K , D^{}_L, \psi^{}_0, \chi^{}_0, \phi^{}_0\}\,.
\end{eqnarray}
The Fisher matrices for EMRI waveforms have very large condition numbers (the ratio of the largest to the smallest 
eigenvalues) and so we use an LU decomposition to invert them, where the Fisher matrix [Eq.~(\ref{covariance})] 
is written as the product of a lower triangular matrix and an upper triangular matrix~\cite{Huerta:2008gb}. 
In addition, to verify that the error estimates that we obtain are reliable, we employ the maximum-mismatch criterion 
proposed in~\cite{Vallis2008}.

%----RESULTS----%
\section{Results}\label{Results}
The main goal of our study is to determine how well a GW detector like LISA could estimate the parameters of an 
EMRI system (see Table~\ref{parameters}) in DCSMG and in particular how well it could constrain the CS parameter $\xi$ 
and hence distinguish between an inspiral in DCSMG and one in GR. To that end, we have followed the approach 
of~\cite{Sopuerta:2009iy} to model the motion of the SCO in the modified DCSMG MBH geometry, but this time we have 
included RR effects.  To evolve the inspiral we have followed the algorithmic scheme described above.   Once the
trajectory of the SCO has been computed, we obtain the waveforms by using the multipolar formalism described by Thorne
in~\cite{Thorne1980R}, but only keeping the quadrupolar term.  To compute the Fisher matrix, the waveform derivatives,
$\partial_i\textbf{h}=\partial\textbf{h}/\partial \lambda^i$, are obtained numerically through a five-point finite 
difference rule. Finally, the response of the detector and the Doppler modulation for a detector like LISA are computed 
using the expressions in~\cite{Barack:2003fp} and in~\cite{Cornish2003}. 
%
%------------------------- Table 1: --------------------------------
\begin{table*}[h!]
\caption{Summary of the EMRI parameter space employed in our analysis. The angles $(\theta^{}_S, \phi^{}_S)$ and 
$(\theta^{}_K, \phi^{}_K)$ are spherical coordinates and $t^{}_0$ stands for the initial time in the computation. 
For the parameters with physical dimensions, we indicate the units used in round brackets. \label{parameters}} 
\centering 
\scalebox{1}{   
\begin{tabular}{l l}
\br\\[-2mm]
$M_{\bullet} $			&MBH mass.\\
$a =|\bold{S}|/M_{\bullet}$~~$(M^2_{\bullet})$& MBH spin.\\
$\mu  = m/M_{\bullet} $ & Mass-ratio.\\
$e^{}_0$			   		& Eccentricity of the particle orbit at $t^{}_0$.\\
$p^{}_0$			   		&Semilatus rectum at $t^{}_0$.\\
$\theta^{\INC}_{0}$  		&Inclination of the orbit at $t^{}_0$.\\
$\hat{\xi} = \xi\times S $~~$(M^{6}_{\bullet})$ &  Product of the Chern-Simons parameter and the spin.\\
$\theta^{}_S$		    & Polar angle with respect to the ecliptic.\\
$\phi^{}_S$		    & Azimuthal angle with respect to the ecliptic.\\
$\theta^{}_K$		    & Spin polar angle with respect to the ecliptic.\\
$\phi^{}_K$		    & Spin azimuthal angle with respect to the ecliptic.\\
$D^{}_L$~~(Gpc)		    & Distance to the EMRI from the ecliptic baricentre.\\
$\psi^{}_0$		    &Phase angle associated with the radial motion at  $t^{}_0$.\\
$\chi^{}_0$		    & Phase angle associated with the polar motion at $t^{}_0$.\\
$\phi^{}_0$ 	   &Boyer-Lindquist azimuthal coordinate, $\phi$, at $t^{}_0$.\\
\br
\end{tabular} }
\end{table*} 
%

%------------------------- Table 2: --------------------------------%
\begin{table*}[h!!]
\caption{Parameters of the different systems used in our work, where we have fixed $p =10$, 
$\theta^{}_{\INC}=0.85$, $\theta^{}_S=1.1$, $\phi^{}_S=0.3$, $\theta^{}_K=1.4$, $\phi^{}_K=0.25$, 
$D^{}_L=1\,$Gpc, $\psi^{}_0=0.25$,  $\chi^{}_0=1$, and $\phi^{}_0=0.1$.}\label{systems}  
\centering  
\begin{tabular}{ l c c c c c}
\br
System & $M_{\bullet}$  & $a/M_{\bullet}$          & $\mu$                     
& $e_0$       & $\hat{\xi}/M_{\bullet}^6$ \\
\hline
A          &  $5\times10^5$ & $0.5$    & $2\times10^{-5}$    & $0.5$  
&$10^{-2}$\\
\hline 
B          &  $5\times10^5$ & $0.5$    & $2\times10^{-5}$    & $0.5$  & $0$    
\\
\hline 
C          &  $5\times10^5$ & $0.25$  & $2\times10^{-5}$    & $0.5$  & $10^{-2}$
\\
\hline
D          &  $5\times10^5$ & $0.25$  & $2\times10^{-5}$    & $0.5$  & $0$      
  \\
\hline
E          &  $10^6$         & $0.25$   & $10^{-5}$               & $0.32$ 
&$10^{-2}$ \\
\hline 
F          &  $10^6$            & $0.25$   & $10^{-5}$               & $0.32$  &
$0$       \\
\br
\end{tabular}
\end{table*} 
%--------------------------------

We have computed parameter measurement error estimates for different systems in DCSMG and the analogous 
systems in GR (that is, setting $\xi =0=\delta\xi$).  In Table~\ref{systems}, we show the parameters of 
the different (systems) used in this analysis, where the distance to the source is given in Gpc, 
the angles are given in radians and the rest of the parameters are given in MBH mass units, while the 
MBH mass itself is given in solar masses. The direction of the MBH spin is taken to be parallel to the 
$z$-direction of the EMRI reference frame.

As a test of our numerical set-up, in Table~\ref{accuracy_1} we compare our numerical estimates in GR with 
the ones obtained in~\cite{Barack:2003fp}. Since the parameter space employed in our analysis differs with 
the one chosen in~\cite{Barack:2003fp}, we just compare the results obtained for the parameters common to  
both works and, as we can see from Table~\ref{accuracy_1}, the two methods produce very similar results. 
However, as it is shown in the table, our estimate for the error in the MBH spin is two orders of magnitude smaller 
than the one obtained in~\cite{Barack:2003fp}. In our analysis, we include the effect of the MBH spin directly in the geodesic equations, whereas in~\cite{Barack:2003fp} the effect of the MBH spin is introduced as a post-Newtonian correction in their {\em post-Keplerian} equations of motion. Our analysis therefore includes  higher order spin corrections which may explain some of the improvement that our results show. A similar improvement in spin measurement precision was observed in~\cite{Huerta:2008gb} who used a similar method based on exact Kerr geodesic trajectories.
%In addition, for these results we are considering only a subspace of the waveform parameters, whereas in~\cite{Barack:2003fp} they computed results for the full fourteen dimensional EMRI parameter space. Our analysis may therefore be ignoring some correlations in the waveform that limit the measurement accuracy for the spin parameter.

%------------------------- Table 3: --------------------------------%
\begin{table*}[h!!]
\caption{Parameter accuracy estimates for the inspiral of a $10 M_{\odot}$ SCO into a $10^6M_{\odot}$ MBH. It 
is assumed that we have observed the last year of inspiral and that the system is an EMRI described by GR, i.e., 
$\xi=0$. We compare the results we have obtained with the ones obtained for the same parameters by
Barack and Cutler~\cite{Barack:2003fp}.\label{accuracy_1} } 
\centering
 \begin{tabular}{l c c}
\br
Parameter				 &GR (this work)		&GR (\cite{Barack:2003fp}) \\
\mr
$\Delta \ln{M_{\bullet}}$	&$8.6\times10^{-4}$	&$9.2\times10^{-4}$\\
$\Delta a/M_{\bullet}$ $(M_{\bullet})$		&$9.5\times10^{-6}$ &$6.3\times10^{-4}$\\
$\Delta \ln{\mu}$ 		&$1.4\times10^{-4}$ &$9.2\times10^{-5}$\\
$\Delta e$			       	&$2.7\times10^{-5}$	&$ 2.8\times10^{-4}$\\
$\Delta \ln{D_L}$$\,{}^{(a)}$ , $\Delta \ln{(\mu/ D_L)}$$\,{}^{(b)}$ & $6.5\times10^{-2}$$\,{}^{(a)}$	&$ 3.7\times10^{-2}$$\,{}^{(b)}$\\
\br
\end{tabular}
\end{table*} 
%

%------------------------- Table 4: --------------------------------%
\begin{table*}[h!!]
\caption{Error estimates for half a year of evolution (SNR$= 35$) for Systems A, B, C, and D; and for one year of evolution 
(SNR$=40$) for Systems E and  F. \label{cs_results}} 
\centering    
\begin{tabular}{l  c  c  c  c c c }
\br
Parameter				& A			    	&B				&C			       &D	  			&E				&F\\
\mr\\[-2mm]
%1
$\Delta\ln{M_{\bullet}}$	&$1.0\times10^{-4}$&$9.8\times 10^{-3}$	&$7.8\times10^{-4}$&$7.6\times10^{-4}$	&$8.6\times10^{-4}$ 	&$8.4\times10^{-4}$\\
%
%2
$\Delta a$~~$(M_{\bullet})$&$1.5\times10^{-5}$&$9.5\times10^{-6}$	&$7.5\times10^{-6}$&$6.3\times10^{-6}$	&$1.8\times10^{-5}$	&$9.1\times10^{-6}$\\
%
%3
$\Delta \ln{\mu}$ 		& $1.8\times10^{-4}$&$3.9\times10^{-5}$	&$5.7\times10^{-5}$&$2.4\times10^{-5}$ 	&$3.2\times10^{-4}$	&$1.2\times10^{-4}$\\
%
%4
$\Delta e_0$			      	 &$9.2\times10^{-5}$&$9.2\times10^{-5}$	&$8.4\times10^{-5}$&$8.4\times10^{-5}$	&$1.9\times10^{-5}$	&$1.7\times10^{-5}$\\
%
%5
$\Delta \theta^{\INC}_{0}$   	&$2.8\times 10^{-5}$&$1.7\times10^{-5}$	&$2.6\times10^{-5}$&$2.2\times10^{-5}$	 &$8.1\times10^{-5}$	&$3.2\times10^{-5}$\\
%
%6
$\Delta\ln{\hat{\xi} }$  	&$2.1$~~~	       &				&$4.4\times10^{-1}$&				&$5.1$		        &\\
%
%7
$\Delta \theta_S$		&$1.8\times10^{-2}$&$1.7\times10^{-2}$	&$1.5\times10^{-2}$&$1.4\times10^{-2}$  	 &$1.1\times10^{-2}$	&$1.1\times10^{-2}$\\
%
%8
$\Delta\phi_S$		    	&$1.7\times10^{-2}$&$1.6\times10^{-2}$ 	&$1.5\times10^{-2}$&$1.5\times10^{-2}$	 &$2.1\times10^{-2}$	&$2.0\times10^{-2}$\\
%
%9
$\Delta\theta_K$		&$2.8\times10^{-2}$&$2.8\times10^{-2}$	&$2.5\times10^{-2}$&$2.5\times10^{-2}$	 &$2.3\times10^{-2}$	&$2.3\times10^{-2}$\\
%
%10
$\Delta\phi_K$		        &$7.2\times10^{-2}$&$7.2\times10^{-2}$     &$6.5\times10^{-2}$&$6.5\times10^{-2}$	 &$8.6\times10^{-2}$	&$8.6\times10^{-2}$\\
%
%11
$\Delta D_L$~~$(Gyr)$  &$2.7\times10^{-2}$&$2.6\times10^{-2}$  	&$2.2\times10^{-2}$&$2.3\times10^{-2}$	 &$1.8\times10^{-2}$	&$1.9\times10^{-2}$\\
%
%12
$\Delta \psi_0$		       &$9.8\times10^{-2}$&$7.1\times10^{-2}$	&$7.2\times10^{-2}$&$6.0\times10^{-2}$	 &$1.0\times10^{-2}$	&$8.6\times10^{-2}$\\
%
%13 
$\Delta \chi_0$		       &$8.6\times10^{-2}$&$8.5\times10^{-2}$  	&$7.7\times10^{-2}$&$7.7\times10^{-2}$	 &$9.2\times10^{-2}$	&$9.0\times10^{-2}$\\
%
%14
$\Delta \varphi_0$ 	      &$8.6\times10^{-2}$&$8.3\times10^{-2}$       &$7.6\times10^{-2}$&$7.5\times10^{-2}$        &$1.1\times10^{-1}$&$1.1\times10^{-2}$\\
\br
\end{tabular} 
\end{table*} 

We have considered two different kind of EMRIs, namely one composed of a MBH with $M_{\bullet} = 5\times10^5M_{\odot}$ 
and a SCO with mass $m=10M_{\odot}$, and the other one composed of a MBH with mass $M_{\bullet} = 10^6M_{\odot}$ and a
SCO with mass $m=10M_{\odot}$.  For the former, we consider an observation lasting for the six months before plunge
(SNR$=35$) and for the latter we consider an observation lasting one year before plunge (SNR$=40$). In Table~\ref{cs_results}, 
we show the error estimates obtained in our simulations.  We typically find that an EMRI observation can determine the 
extrinsic parameters, namely $\vec{\lambda}_{\rm extrinsic}= \{\theta^{}_S, \phi^{}_S, \theta^{}_K, \phi^{}_K ,D^{}_L, 
\psi^{}_0, \chi^{}_0, \phi^{}_0\}\,$, to a fractional error of $\sim 10^{-2}$, while the intrinsic parameters: 
$\vec{\lambda}_{\rm intrinsic}= \{M_{\bullet}, a, \mu, e^{}_0, \theta^{\INC}_0\}$, are typically measured with a fractional 
error of $\sim10^{-4}, 10^{-5}, 10^{-5}, 10^{-5},10^{-5}$ respectively. For the CS parameter $\hat{\xi}$, we find that 
in the best case it can be measured with a fractional error $\sim10^{-1}$. These results suggest that LISA might be able to distinguish between GR and DCSMG using certain EMRI observations. Notice that in this paper we are using the concept of the LISA mission as a reference and, as it is shown in~\cite{cgs:2012}, the results obtained are also valid for the updated eLISA or NGO (New Gravitational-Waves Observatory)~\cite{AmaroSeoane:2012km} by normalising to a fixed SNR. In addition, we find that for a given value of $\hat{\xi}$, it will be better estimated for systems with smaller MBH mass, $M_{\bullet}$, and spin, $a$. This is not 
surprising, since the leading-order term that contains the CS parameter in the DCSMG metric [Eq.~(\ref{DCSMG_metric})] is 
$\sim \hat{\xi} /M^6_{\bullet} (M_\bullet/r)^6$. By decreasing the MBH mass with fixed $\hat{\xi}$, this term becomes larger at fixed $r/M_\bullet$ and by decreasing the MBH spin this term is larger relative to the Kerr $g_{t\phi}$ metric coefficient. 

Due to the relatively high computational cost of these calculations, a full (i.e., including all the parameters) error 
estimation takes approximately three days using a single processor machine. For this reason, we have only studied a small 
number of points in the parameter space of all possible EMRI systems that we might detect. In this sense, these results could 
be improved by performing a more exhaustive study of the parameter space and also using a full Monte Carlo sampling. 
In addition, we could have considered longer gravitational waveforms, that cover the entire length of the LISA mission 
($\sim2$--$5$ years), which would be expected to improve our ability to measure the system parameters.

%----- CONCLUSIONS-----%
\section{Conclusions}
\label{conclusions}

In this work we have presented a first study aimed at assessing the ability of a GW detector like LISA to discriminate 
between GR and DCSMG by performing parameter estimation studies of EMRIs in DCSMG.  We have computed the waveforms 
generated by a SCO orbiting in a MBH geometry modified by CS corrections. The parameter error estimates have been 
computed using a Fisher matrix analysis. To validate our computational framework, we have studied a typical EMRI system 
in GR and we have found agreement between our results and previous results found in the literature. We have applied these 
parameter estimation studies to estimate the ability of LISA to distinguish between GR and DCSMG by estimating the 
CS parameter $\xi$. To that end, we have performed simulations of the GWs emitted during the last six months before
plunge of an EMRI system for which the gravitational wave emission is at frequencies in the most sensitive part of the 
LISA band. Our results indicate that, for certain EMRI systems, a detector like LISA may be able to discriminate between 
GR and DCSMG. We have also seen that the error in estimating $\xi$ decreases with the MBH mass and spin. We are now 
performing a more exhaustive study of the parameter space of possible EMRIs to better understand if these results hold 
in general~\cite{cgs:2012}.

Another subject that deserves further study is the question of the systematic errors that would arise by using GR 
waveform templates to detect EMRIs that are actually described by DCSMG. This can be done by using the formalism 
developed recently by Cutler and Vallisneri~\cite{Cutler:2007}, which allows for the estimation of the magnitude of the 
model errors. Finally, the present work can be extended to other GW detectors, for instance, to study Intermediate-Mass-Ratio
Inspiral detection with the future Einstein Telescope.

%----- ACKNOWLEDGMENTS-----%

\section*{Acknowledgments}
We would like to thank Michele Vallisneri for helpful discussions. 
PCM is supported by a predoctoral FPU fellowship of the Spanish Ministry of Science and Innovation (MICINN). 
JG's work is supported by the Royal Society.
CFS acknowledges support from the Ram\'on y Cajal Programme of the Ministry of Education and Science of Spain, 
from a Marie Curie International Reintegration Grant (MIRG-CT-2007-205005/PHY) of the 7th European Community 
Framework Programme, from contract AYA-2010-15709 of the MICINN, 
and from contract 2009-SGR-935 of the Catalan Agency for Research Funding (AGAUR).   We acknowledge the 
computational resources provided by the Barcelona Supercomputing Centre (AECT-2011-3-0007) and CESGA 
(contract CESGA-ICTS-200).

%----- APPENDIX -----%
\appendix
%

%----- BIBLIOGRAPHY-----%
\section*{References}

%\bibliography{References}

%----- END-----%
\end{document}